\documentclass[unsortedaddress,
twocolumn,showpacs,preprintnumbers,amsmath,amssymb]{revtex4}
\usepackage{graphicx}
\usepackage{dcolumn}
\usepackage{bm}
\usepackage{latexsym}
\begin{document}
\title{Collective traffic-like movement of ants on a trail:\\
dynamical phases and phase transitions
}
\author{Ambarish Kunwar}
\email{ambarish@iitk.ac.in} 
\affiliation{%
Department of Physics, Indian Institute of Technology,
Kanpur 208016, India.
}%
\author{Alexander John}%
 \email{aj@thp.uni-koeln.de}
\affiliation{%
Institut  f\"ur Theoretische  Physik, Universit\"at 
zu K\"oln D-50937 K\"oln, Germany
}%
\author{Katsuhiro Nishinari}
\email{knishi@rins.ryukoku.ac.jp}
\affiliation{%
Department of Applied Mathematics and Informatics, 
Ryukoku University, Shiga, Japan 
}%
\author{Andreas Schadschneider}%
 \email{as@thp.uni-koeln.de}
\affiliation{%
Institut  f\"ur Theoretische  Physik, Universit\"at 
zu K\"oln D-50937 K\"oln, Germany
}%
\author{Debashish Chowdhury}
\email{debch@iitk.ac.in} 
\affiliation{%
Department of Physics, Indian Institute of Technology,
Kanpur 208016, India.
}%
\date{\today}%
\begin{abstract}
The traffic-like collective movement of ants on a trail can be
described by a stochastic cellular automaton model. We have
earlier investigated its unusual flow-density relation by using
various mean field approximations and computer simulations.
In this paper, we study the model following an alternative approach
based on the analogy with the zero range process, which is one of the
few known exactly solvable stochastic dynamical models. We show that
our theory can quantitatively account for the unusual non-monotonic
dependence of the average speed of the ants on their density for finite
lattices with periodic boundary conditions. Moreover, we argue that the
model exhibits a continuous phase transition at the critial density only
in a limiting case. Furthermore, we investigate the phase diagram of the
model by replacing the periodic boundary conditions by open boundary
conditions.
\end{abstract}
\maketitle
\section{Introduction}
It is well known that no phase transition takes place in one-dimensional 
systems in thermodynamic equilibrium if the interactions are short-ranged. 
But, systems of interacting particles driven far from equilibrium can 
exhibit transitions from one dynamical phase to another even in 
one-dimensional space with only short-ranged interactions \cite{sz,schutz}. 
A class of models, which have been receiving lot of attention in the recent 
years from this perspective, consist of interacting particles each of 
which is self-propelled rather than driven by external field 
\cite{css}; the dynamics of these systems are most often formulated in 
terms of updating rules and, therefore, these models can also be 
regarded as stochastic cellular automata (CA) \cite{wolfram,droz}.

Several different mechanisms are known to give rise to the phase 
transitions in systems of driven interacting particles. The 
boundary-induced phase transitions \cite{krug} have been studied 
extensively \cite{schutz}. Phase transition has also been observed 
in a special limiting situation of the bus route model \cite{LEC} 
that involves two coupled dynamical variables one of which, because 
of the periodic boundary conditions, is conserved while the other is 
not. The aim of this paper is to investigate the dynamical phases and 
phase transitions in a model of interacting {\it self-driven} particles 
\cite{cgns,nds} which has been motivated by the collective traffic-like 
movement of ants on a trail \cite{burd}.

In this paper we first utilize the close relation between the ant-trail 
model (ATM) and zero-range processes (ZRP) \cite{spi,brazil} to 
show the existence of a continuous phase transition
from an inhomogeneous jammed
   phase to homogeneous congested phase
 at a particular 
density in a special limit of this ATM with periodic boundary conditions; 
this phenomenon is argued to be closely related to a phase transition 
in the bus route model \cite{LEC}. Then, in order to investigate the 
boundary-induced phase transitions in the ATM we also compute its phase 
diagram with open boundary conditions for a wide range of values of the 
relevant parameters.

The paper is organized as follows: the ATM is defined in Section 
\ref{sec2} where the corresponding numerical results are summarized 
briefly. Then we discuss the relation between ATM and ZRP in Section 
\ref{sec3} where we also give analytical calculations based on this 
analogy, including the results in the thermodynamical limit. The 
phase diagram of the ATM for open boundary conditions is given in 
Section \ref{sec4}. Conclusions are drawn in Section \ref{sec5}.

\section{Ant trail model}\label{sec2}

Since our ATM model reduces to the asymmetric simple exclusion process 
(ASEP) in some special limits, we begin by defining the ASEP \cite{ASEP}.
A fraction of the sites of a one-dimensional lattice are occupied 
initially randomly by particles each of which can move forward by one 
lattice spacing, with probability $q_{\text{eff}}$, if and only if the target 
site is not already occupied by another particle. The updating is done 
either in parallel or in a random-sequential manner. The ASEP \cite{ASEP} 
with parallel updating has been used often as an extremely simple model 
of vehicular traffic on single-lane highways \cite{css}. 

\subsection{Definition of the ATM model}

Let us now define ATM which is a simple model for an unidirectional
motion of ants on an existing trail. The ants  communicate with each 
other by dropping a chemical (generically called {\it pheromone}) on 
the substrate as they crawl forward \cite{camazine,mc}. 
The pheromone sticks to the substrate long enough for the other 
following sniffing ants to pick up its smell and follow the trail. 
The presence of pheromone leads to a higher effective speed of the 
isolated ants. In ref.8 and 9 we have proposed an extension of the 
ASEP that takes into account this enhancement of the effective speed.

Each site of our one-dimensional ant-trail model represents a cell
that can accomodate at most one ant at a time.
The lattice sites are labelled by the index $i$ ($i = 1,2,...,L$);
$L$ being the length of the lattice. We associate two binary variables
$S_i$ and $\sigma_i$ with each site $i$ where $S_i$ takes the value
$0$ or $1$ depending on whether the cell is empty or occupied by an ant.
Similarly, $\sigma_i =  1$ if the cell $i$ contains pheromone; otherwise,
$\sigma_i =  0$. 
The instantaneous state (i.e., the configuration) of the system at
any time is specified completely by the set $(\{S\},\{\sigma\})$.

Since a unidirectional motion is assumed, ants do not move backward.
Their forward-hopping probability is higher if it smells pheromone 
ahead of it.
The state of the system is updated at each time step in {\it two
stages}.  In stage I ants are allowed to move. This motion follows 
rules similar to those of the particles in the ASEP except that 
the hopping probability now depends on the presence or absence of 
pheromone at the target site. Here the subset
$\{S(t+1)\}$ at the time step $t+1$ is obtained using the full information
$(\{S(t)\},\{\sigma(t)\})$ at time $t$. Stage II corresponds to the
evaporation of pheromone. Here only the subset $\{\sigma(t)\}$ is
updated so that at the end of stage II the new configuration
$(\{S(t+1)\},\{\sigma(t+1)\})$ at time $t+1$ is obtained.
In each stage the dynamical rules are applied {\it in parallel} to
all ants and pheromones, respectively.\\

\noindent {\it Stage I: Motion of ants}\\[0.2cm]
\noindent An ant in cell $i$ that has an empty cell in front of it, i.e., 
$S_i(t)=1$ and $S_{i+1}(t)=0$, hops forward with
\begin{equation}
{\rm probability} = \left\{\begin{array}{lll}
            Q &\quad{\rm if\ }~\sigma_{i+1}(t) = 1,\\
            q &\quad{\rm if\ }~\sigma_{i+1}(t) = 0,
\end{array} \right.
\end{equation}
where, to be consistent with real ant-trails, we assume $ q < Q$.\\

\noindent {\it Stage II: Evaporation of pheromones}\\[0.2cm]
\noindent At each cell $i$ occupied by an ant after stage I
a pheromone will be created, i.e., 
\begin{equation}
\sigma_i(t+1) = 1\quad {\rm if\ }\quad S_i(t+1) = 1.
\end{equation}
On the other hand, any `free' pheromone at a site $i$ not occupied
by an ant will evaporate with the probability $f$ 
per unit time, i.e., if $S_i(t+1) = 0$, $\sigma_i(t) = 1$, then
\begin{equation}
\sigma_i(t+1) = \left\{\begin{array}{lll}
0 &\quad {\rm with\ probability\ } f,\\
1 &\quad {\rm with\ probability\ } 1-f.
\end{array} \right.
\end{equation}

Note that, if periodic boundary conditions are imposed, the dynamics 
conserves the number $N$ of ants, but not the number of pheromones;  
in that case ATM model is a stochastic CA model with two coupled 
dynamical variables one of which is conserved and the other nonconserved. 
The stationary states of this ATM model reduces to those of the ASEP 
\cite{ASEP} with $q_{\text{eff}} = Q$ and $q_{\text{eff}} = q$, 
respectively, in limiting cases $f = 0$ and $f = 1$. 

\subsection{Relation between ATM and the bus route model} 

The bus route model \cite{LEC} describes a system of buses that 
move unidirectionally from one bus stop to the next on a circular 
route and, at each bus stop, a bus picks up the waiting passengers 
that arrive stochastically since the departure of the last bus 
from that stop.

In the bus route model \cite{LEC} the bus stops are represented 
by the sites on a one-dimensional lattice each of which may be 
labeled by an index $i$ ($i = 1,2,...,L$). Two binary variables 
$\sigma_i$ and $\tau_i$ are assigned to each cell $i$: 
(i) If the cell $i$ is occupied by a bus then $\sigma_i = 1$; 
otherwise $\sigma_i = 0$. (ii) If cell $i$ has passengers waiting 
for a bus then $\tau_i = 1$; otherwise $\tau_i = 0$. Since a cell 
cannot have simultaneously a bus and waiting passengers, a cell
cannot have both $\sigma_i = 1$ and $\tau_i = 1$ simultaneously.
Each bus is assumed to hop from one stop to the next.

The {\it random-sequential update} rules of the model are as follows: 
a cell $i$ is picked up at random. Then, (i) if $\sigma_i = 0$ 
and $\tau_i = 0$ (i.e, cell $i$ contains neither a bus nor waiting 
passengers), then $\tau_i \rightarrow 1$ with probability $\lambda$, 
where $\lambda$ is the probability of arrival of passenger(s) at the 
bus stop.  (ii) If $\sigma_i = 1$ (i.e., there is a bus at the cell 
$i$) and $\sigma_{i+1} = 0$, then the hopping rate $\mu$ of the bus
is defined as follows: (a) if $\tau_{i+1} = 0$, then
$\mu = \alpha_b$ but (b) if $\tau_{i+1} = 1$, then $\mu = \beta_b$,
where $\alpha_b$ is the hopping rate of a bus onto a stop which
has no waiting passengers and $\beta_b$ is the hopping rate onto
a stop with waiting passenger(s). Generally, $\beta_b < \alpha_b$,
which reflects the fact that a bus has to slow down when it
has to pick up passengers. One can set $\alpha_b = 1$ without
loss of generality. When a bus hops onto a stop $i$ with waiting
passengers $\tau_i$ is reset to zero as the bus takes all the
passengers. Note that, because of the periodic boundary conditions, 
the density of buses is a conserved quantity whereas that of the 
passengers is not.

Note that in the ATM the pheromones ``appear'' at a site when an ant 
visits the site just as the waiting passengers ``disappear'' 
from a site when a bus visits the site. Also note that the 
``disappearance'' of pheromone from a site takes place with 
a probability $f$ per unit time, independent of the ants. 
Similarly, the passengers ``appear'' at a bus stop with the 
probability $\lambda$ per unit time independent of buses. 
Therefore, the ``appearance'' of the pheromone is analogous 
to the ``disappearance'' of the passengers and vice versa. 
Naturally, we expect the role of the parameter $f$ in the  
ATM to be similar to that of $\lambda$ in the bus route model.
The presence of passenger is analogous to the absence of pheromone. 
Thus, the ATM is equivalent to the BRM with {\it parallel updating} 
\cite{chdesai}.

Interestingly, the queueing of the buses in the bus route model, 
in turn, is very similar to the bunching of macrosteps during 
the growth of faceted crystals because of impurity absorption 
\cite{vander}. If $y_n(t)$ denotes the position of the particle 
(representing a bus or an ant or a macrostep) at time $t$, and 
$\tau_n$ is the time elapsed since the particle ahead (which is 
labelled by $n+1$) passed the same position $y_n(t)$, one can 
write \cite{vander} 
$y_n(t) = y_{n+1}(t - \tau_n)$ and corresponding speed 
$v_n(t) = V(\tau_n(t))$ is a function of $\tau_n(t)$.

\subsection{Numerical results}

We shall use the symbols $L$ and $M$ to denote the system size and 
the number of ants, respectively; hence $\rho=M/L$ is the density 
of ants on the trail. 
The most important quantity of interest in the context of flow 
properties of the traffic models is the {\it fundamental diagram}, 
i.e., the flux-versus-density relation, where flux $F$ is the product 
of the density $\rho$ and the average speed $v$. Because of the similarity 
of the ATM with traffic models, it would be interesting to draw 
the fundamental diagram of ATM and compare the traffic-like 
collective movements of ants with vehicular traffic.

The fundamental diagrams of ATM is given in our earlier papers 
(see, for example, Fig.2(b) of ref.8 or Fig.2(b) of ref.9).
First of all, the diagram does not possess particle-hole symmetry 
for any $f$ in the range $0 < f < 1$; the particle-hole symmetry 
observed in the ATM in the special cases of $f = 0$ and $f = 1$  
is a consequence of the fact that in the two special cases $f = 0$ 
and $f = 1$, as pointed out earlier, the ATM becomes identical to 
the ASEP, with parallel updating, corresponding to the effective 
hopping probabilities $q_{\text{eff}} = Q$ and $q_{\text{eff}} = q$, respectively. 

Most important feature of the fundamental diagram is that, over a 
range of small values of $f$ ($f \lesssim 0.01$), it exhibits an 
anomalous behaviour; this is a consequence of the fact that, unlike 
common vehicular traffic, the average speed of the ants in the ATM 
is a non-monotonic function of the density of ants on the trail 
(see, for example, Fig.3(b) of ref.8 or Fig.2(a) of ref.9).

The ATM model also exhibits interesting coarsening of clusters of 
ants starting from random initial conditions (Fig.1). 
\begin{figure}[ht]
\begin{center}
\includegraphics[width=0.7\columnwidth]{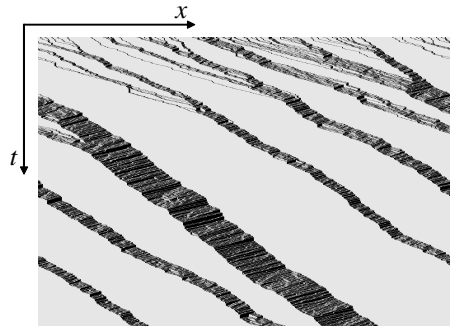}
\caption{Spatial-temporal behaviours of loose clusters 
in the low density case ($\rho=0.16$). Parameters
are $Q = 0.75, q = 0.25, f=0.005$). We see the loose clusters emerge
 form the random initial configuration, and will eventually merge
into one big loose cluster if we take sufficiently long time.}
\end{center}
\end{figure}

\section{Phase transition in ATM with periodic boundary conditions}\label{sec3}
In ref.9 we have developed a formalism by introducing the loose 
cluster approximation and successfully captured the non-monotonicity 
heuristically with the help of analytical results of ASEP, which is 
an exactly solvable stochastic model. In this paper, we study the 
phenomenon in detail by utilizing the analogy of ATM with the zero 
range process (ZRP), which is exactly solvable \cite{spi,brazil}.
The ZRP is a particle-hopping model where the hopping probabilities 
do not depend on the state of occupation of the target site.

\subsection{Analytical results for finite systems: ZRP and ATM}
We now first show the relation between the ATM and the ZRP 
\cite{spi,brazil}. Note that, in the ATM the effective hopping 
probability $u$ of the can be exactly expressed as \cite{cgns,nds}
\begin{equation}
 u=q(1-g) + Q g,
\label{hop}
\end{equation}
where $g$ is the probability that there is a surviving pheromone 
on the first empty site of a gap in front of an ant. Obviously, 
$g(t)$ depends on the time $t$ elapsed since the ant immediately 
in front left the target site. It is straightforward to see that 
\begin{equation}
g(t+1)=(1-f)g(t) 
\label{gtime}
\end{equation}

Alternatively, we can express the probability $g(t)$ as a function 
$\tilde{g}(x)$ of the gap $x$ between the ant under consideration 
and that immediately in front of it. Recall that the average 
speed of the ants is $v$.  Since the time interval between the 
passage of successive ants through any arbitrary site is $x/v$, 
and since equation (\ref{gtime}) holds for all $t$, we obtain 
\begin{equation}
\tilde{g}(x)=(1-f)^{x/v}  
\label{gx}
\end{equation}
after iterating the mapping (\ref{gtime}) $x/v$ times. 

Thus, the ATM can be regarded as an ASEP involving fictitious particles 
where, $u(x)$, the effective hopping probability of a particle having 
a gap $x$ in front, depends on $x$ through the relation \cite{cgns}
\begin{equation}
u(x)=q+(Q-q)(1-f)^{x/v}.
\label{hop2} 
\end{equation}
Note that in the case of the original ASEP, $u(x) =$ constant for 
$x > 0$ and $u(0) = 0$. The form (\ref{hop2}) is approximate 
description of the hoppings of the ants in the ATM, but all the 
results we derive for this form of $u(x)$ are exact.	 

As long as $f$ remains non-zero, $u(x)$ decreases with increasing 
$x$ at a rate much faster than that required for a phase transition 
to occur \cite{mohanty}. However, since $v$ depends on $f$ and $x$, 
the trend of variation of $u(x)$ with $x$ is difficult to infer 
without going through a detailed analysis which we present below.

The configurations of this ASEP can be uniquely described by the 
gap configurations $\{x_1,x_2,...,x_M\}$. Note that 
in this ASEP the pheromones do not appear explicitly, but their 
effect enters through the $f$-dependence of $u(x)$.
It is well known \cite{brazil} that any ASEP, where the hopping 
probability of the particles $u(x)$ depends on the gap $x$ in front 
of the particle, can be  mapped onto the ZRP. The advantage of 
mapping the ATM onto the corresponding ZRP (with parallel updating) 
is that the the stationary state of the ZRP is given by product 
measure in spite of the inter-particle interactions. 

In order to avoid any possibility of confusion, we now point out 
the commonly used notation for ZRP \cite{evans04}. In that 
notation for ZRP, the lattice consists of $M$ sites each of 
which can hold an integer number of indistinguishable particles. 
The configuration of the system is specified by the occupation 
numbers $\{n_1,n_2,...,n_M\}$. The total number of particles is 
$L$. On mapping this ZRP onto the ASEP, each site of ZRP becomes a 
particle in ASEP while the total number of sites of the ASEP 
becomes identical to the total number of particles in ZRP. 
Therefore, the expressions of some quantities like, for example, 
the average speed may appear slightly different in the two 
notations, but are completely equivalent.

Following the treatment of the bus route model in ref.7, now 
the steady-state probability $P_M(\{x_{\mu}\})$ of finding the ATM 
in a gap configuration $\{x_1,x_2,...,x_M\}$ is given by a product 
of factors $h(x_{\mu})$, i.e., 
\begin{equation}
P_M(\{x_{\mu}\}) = \frac{\prod_{\mu=1}^M h(x_{\mu})}{Z(L,M)}
\label{stst}
\end{equation}
where the partition function $Z(L,M)$ is given by the relation 
\begin{equation}
Z(L,M) = \sum_{x_1,x_2,...,x_M} \prod_{\mu=1}^M h(x_{\mu}) \delta\biggl(\sum_{\mu} x_{\mu} - (L-M)\biggr).
\label{pfunction}
\end{equation}
The Kronecker detla in (\ref{pfunction}) has the standard meaning, 
i.e., $\delta (a,b)=1$ if $a=b$ and $\delta (a,b)=0$ otherwise.
Note that $Z(L,M)$ is just a normalization which ensures that the 
sum of the probabilities $P_M(\{x_{\mu}\})$ over all possible gap 
configurations $\{x_1,x_2,...,x_M\}$ is unity.

The single-gap probability $p(x)$, i.e., the probability that 
there is a gap of size $x$ in front of an ant, is obtained from 
the $M$-gap probability $P(\{x_{\mu}\})$ by using 
\begin{eqnarray}
&&p(x) = P_1(x) \nonumber \\
&=& 
\frac{\displaystyle{\sum_{x_2,...,x_M}}\delta\biggl(x_2+...+x_M - 
(L-M-x)\biggr) h(x) \prod_{\mu=2}^M h(x_{\mu})}{Z(L,M)} \nonumber \\
&=& h(x) \frac{Z(L-x-1,M-1)}{Z(L,M)}
\label{ph}
\end{eqnarray}
where $Z(L-x-1,M-1)$ is the partition function for a system from 
which the first site has been removed. 

Note that $Z(x,1)=h(x-1)$ and $Z(x,x)=h(0)$. Then, in principle, 
if $h(x)$ is known, the partition function $Z$ can be obtained 
by using the recursion relation
\begin{equation}
 Z(L,M)=\sum_{x=0}^{L-M}Z(L-x-1, M-1)h(x),
\end{equation}

The ATM is formulated with parallel update. The corresponding form 
of $h(x)$ is known to be given by \cite{Evans97}
\begin{equation}
 h(x)=\left\{\begin{array}{cc}
1-u(1)~~~~~~~~ & {\rm for} \,\,\,\,\,\,\, x=0,\\
\displaystyle{\frac{1-u(1)}{1-u(x)}\prod_{y=1}^x \frac{1-u(y)}{u(y)}}~~~~~~~~
& {\rm for} \,\,\,\,\,\,\, x>0
\end{array}
\right.
\label{hform}
\end{equation}

Our aim is to calculate the average velocity $v$ of ants given by
\begin{equation}
 v=\sum_{x=1}^{L-M}u(x)p(x).
\label{veldef}
\end{equation}
In order to compute $v$ we need to compute $p(x)$ and $u(x)$. 
However, the computation of $p(x)$ requires $h(x)$ which, in turn, 
depends on $u(x)$. On the other hand, $u(x)$ is given by equation 
(\ref{hop2}) which involves average speed $v$. Therefore, in order 
to compute $v$ self-consistently, we begin with the initial 
approximate estimate $v = q$ and, hence, compute $u(x)$ and 
$p(x)$ to get better estimate of average speed $v$ from equation 
(\ref{veldef}). Through this iterative process, we estimate $v$ 
self-consistently and, hence, the fundamental diagram.


Fundamental diagrams are given in Fig.2 with $L=100$ 
and $L=200$. The data points, shown with various symbols in 
Fig.2 and connected by dashed curves, have been obtained 
from computer simulations of the ATM, whereas the continuous curves   
are the corresponding theoretical predictions. We find that the 
theoretical curves are almost identical to the numerical ones, thus 
confirming that the ZRP successfully describes the steady state of 
the ATM.

\begin{figure}[ht]
\begin{center}
\includegraphics[angle=-90,width=0.45\columnwidth]{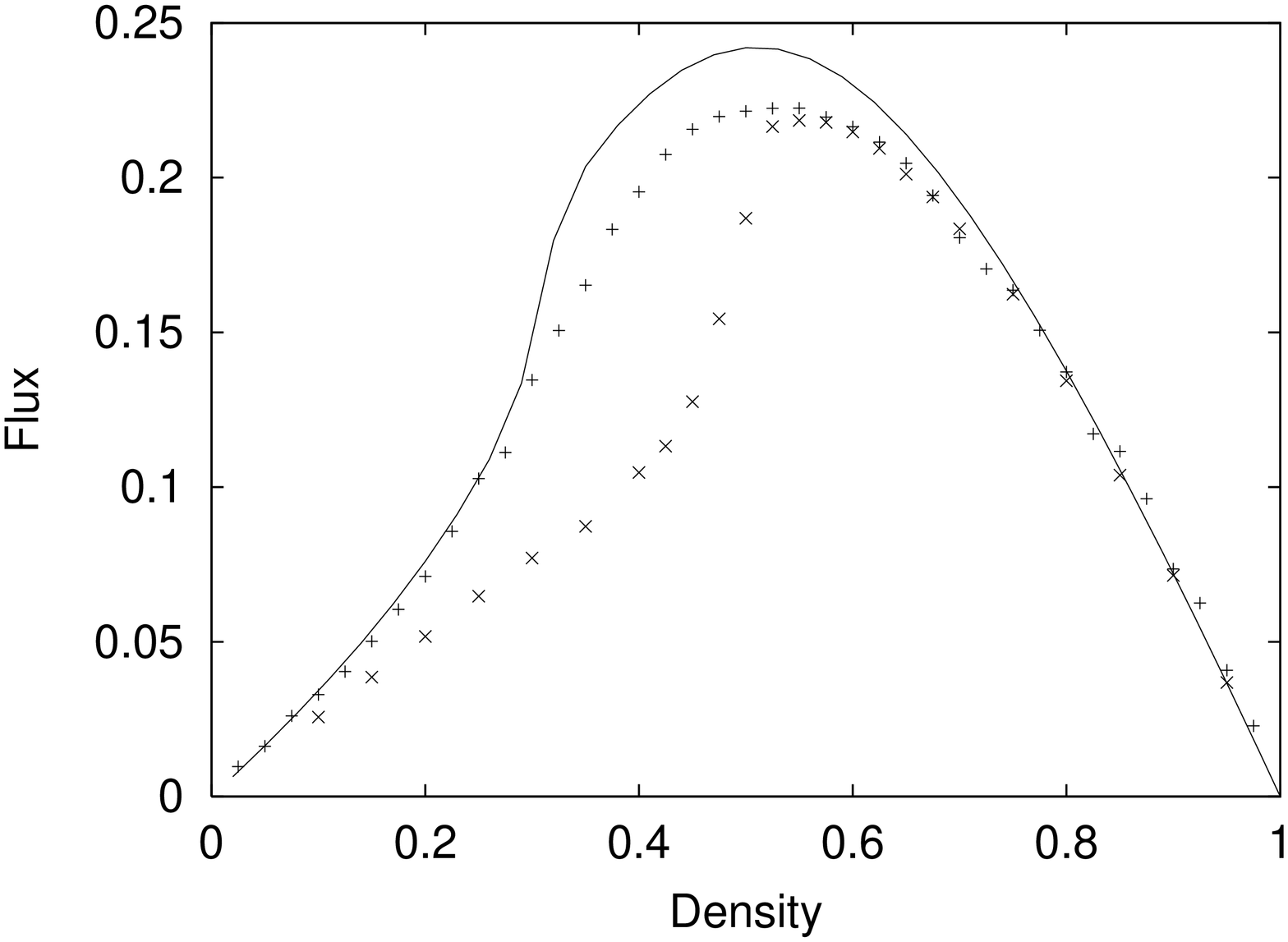}
\includegraphics[angle=-90,width=0.45\columnwidth]{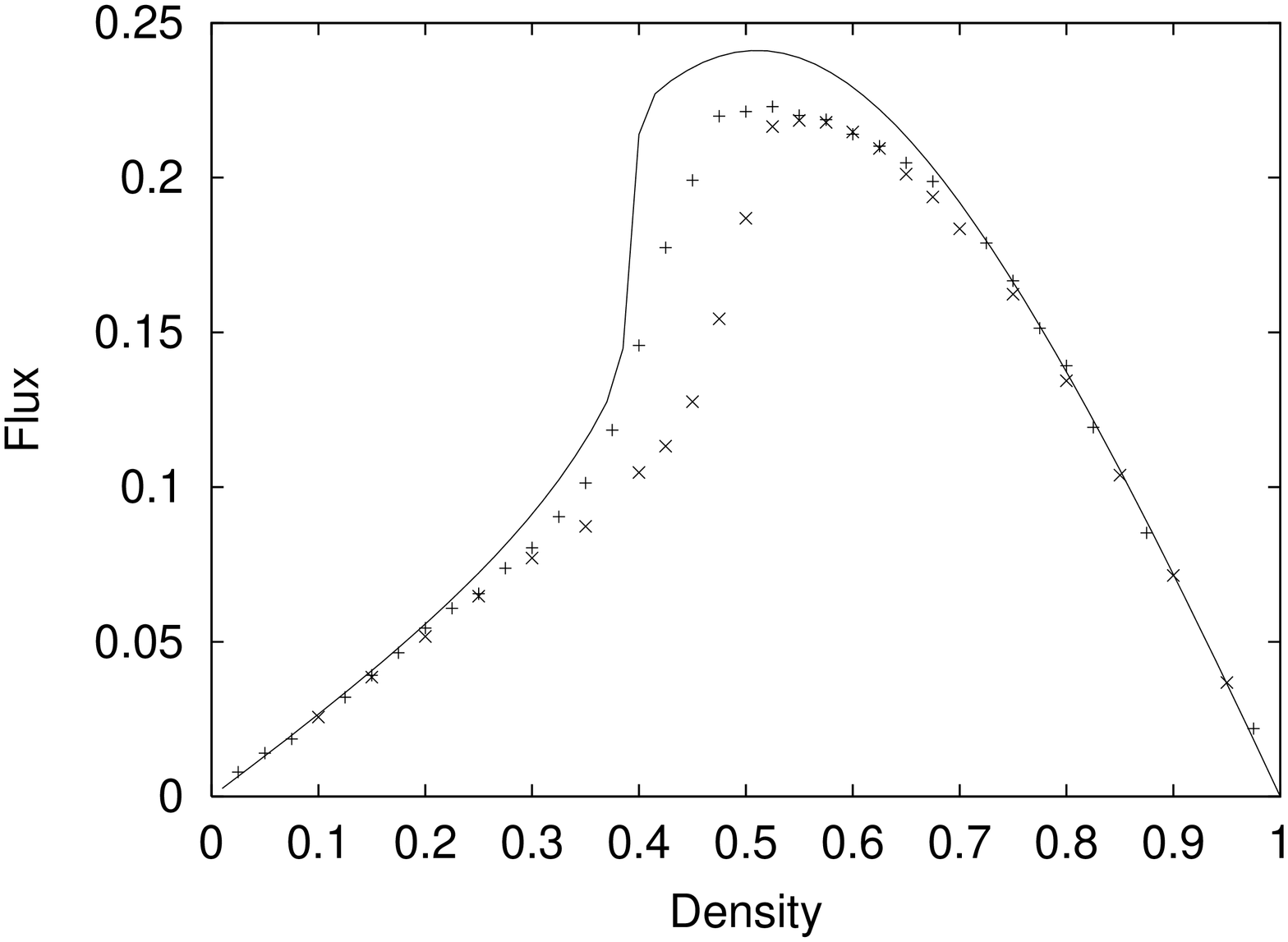}\\
\vspace{0.25cm}
(a)\hspace{3cm}(b)
\caption{Comparison of the theoretically obtained fundamental diagrams 
of the ATM with for the system sizes (a) $L=100$ (continuous curve in 
(a)) and (b) $L=200$ (continuous curve in (b)) with the corresponding 
numerical data obtained from computer simulations (represented by the 
symbol $+$). The numerical data obtained from simulations of systems 
of size $L = 1000$ are also plotted (represented by the symbol $\times$) 
in both (a) and (b) to show the trend of variation with increasing $L$.
The common parameters are $Q=0.75, q=0.25,f=0.005$.
}
\end{center}
\label{zrpfund}
\end{figure}

\subsection{Thermodynamic limit of ATM}
Next we discuss the thermodynamic limit of the ATM, that is,
the case $L \to \infty$.  From Fig.2 we see that the 
curve shows sharp increase near the density region $0.4 <\rho <0.5$, 
and the tendency is expected to increase with the increase of $L$.
Thus it is important to study whether there is a (second order) 
phase transition 
from an inhomogeneous jammed
phase, in which the ants are bunched together, 
to homogeneous congested phase 
with the increase of the density of the ants 
in the thermodynamic limit.

Using the integral representation 
\begin{equation}
\delta(\mu,\nu) = \oint \frac{ds}{2\pi i} \frac{s^{\mu}}{s^{\nu+1}}
\label{kronecker}
\end{equation}
of the Kronecker delta, we rewrite the partition function $Z(L,M)$ as 
\begin{equation}
 Z(L,M)=\oint \frac{ds}{2\pi i s}\left(\frac{G(s)}
{s^{1/\rho-1}}\right)^{M},
\label{pfuncint}
\end{equation}
where $G(s)$ is the generating function of $h$ defined by
\begin{equation}
 G(s)=\sum_{x=0}^{\infty}h(x)s^x. 
\label{gendef}
\end{equation}
We evaluate the integral in (\ref{pfuncint}) in the $L \to \infty$ 
limit, keeping $M/L=\rho$ as constant, by the saddle point method. 
The partition function reduces to the form  
\begin{equation}
 Z(L,M)\sim \exp(M \ln G(z)-(L-M)\ln z)
\label{pfuncsad}
\end{equation}
where the saddle point $s=z$ is given by
\begin{equation}
 \frac{1}{\rho}-1=z\frac{\partial \ln G(z)}{\partial z}.
\label{rho} 
\end{equation}
The equation (\ref{rho}) may also be regarded as the relation that 
defines $z$.

Using the approximate form (\ref{pfuncsad}) of the partition 
function, we have 
\begin{equation}
p(x) \simeq h(x) \frac{Z(L-x-1,M-1)}{Z(L,M)} \simeq \frac{h(x)}{G(z)} z^x. 
\label{pform}
\end{equation}
Then, substituting (\ref{pform}) in expression (\ref{veldef}) we get  
\begin{equation}
 v=\sum_{x=1}^{\infty}\frac{u(x)h(x)}{G(z)}z^x.
\label{vel}
\end{equation}
for the average speed $v$ in the thermodynamic limit. 


Next we study the properties of the generating function $G(z)$ in 
detail to examine the possibility of phase transition of the ATM.
Since $\lim_{x \to \infty} u(x) = q$ for all $f > 0$, the condition 
that $G$ converges is given by
\begin{equation}
 \lim_{x\to\infty}\sup\frac{h(x+1)z}{h(x)} < 1.
\end{equation}
Using (\ref{hform}) for $h(x)$ the convergence condition reduces to 
\begin{equation}
 \lim_{x\to\infty}\sup\frac{h(x+1)z}{h(x)}=\frac{1-q}{q}z < 1.
\end{equation}
Thus, $G$ converges in the range
\begin{equation}
 0<z<z_c=\frac{q}{1-q}.
\label{zcrit}
\end{equation}
Using (\ref{hform}) and (\ref{zcrit}) in (\ref{gendef}) we find that 
the critical value of $G$ at $z=z_c$ is given by
\begin{equation}
 G(z_c)=1-u(1)+\sum_{x=1}^{\infty}\frac{1-u(1)}{1-u(x)}\prod_{y=1}^x
\left(\frac{1-u(y)}{u(y)}\frac{q}{1-q}  \right).
\end{equation}

For all $f>0$, this sum diverges for the class of $u(x)$ which  
decays to $q$ more rapidly than the function $(1+c/x)q$, where 
$c>1-q$. In the ATM, $u(x)=q+(Q-q)(1-f)^{x/v}$, which decays 
exponentially to $q$ as $x \to \infty$. Hence we conclude that 
there is no phase transition in the ATM for $f>0$. This is 
because, from (\ref{rho}), we have $\rho=1$ when $z=0$, and 
$\rho=0$ at $z=z_c$ if $G(z_c)$ diverges. {\it Thus, as long as 
$f > 0$, in the entire density region $0 \le \rho \le 1$ there is 
no singularity in $G$ and, hence, no phase transition in the ATM}.

The situation is drastically different in the limit $f \to 0$.
In this limit, $u(x)=Q$, and then $G(z_c)$ approaches the finite 
limit
\begin{equation}
 \lim_{z \to z_c^{-}}G(z)=\frac{Q(1-q)}{Q(1-q)-q(1-Q)}-Q.
\label{criG}
\end{equation}
Thus, in the limit $f \rightarrow 0$ a phase transition takes place 
at the critical density 
\begin{equation}
 \rho_c=\frac{Q-q}{Q-q^2} 
\label{cri}
\end{equation}
which is obtained from (\ref{rho}). 

In order to get insight into the nature of the phases involved, 
we now calculate the average speed $v = v_c$ at $z = z_c$. 
Since $u(x)\to Q$ as $f \to 0$, $h(x) \to (1-Q)/Q$ and, hence, 
from (\ref{vel}) we get 
\begin{equation}
 v_c=\sum_{x=1}^{\infty}\frac{Q}{G(z_c)}\biggl(\frac{1-Q}{Q}\biggr)^x z_c^x.
\label{criV}
\end{equation}
Substituting (\ref{criG}) into (\ref{criV}), we obtain $v_c=q$.

At first sight the two facts, namely, $u(x) = Q$ and $v_c = q$ 
may appear mutually contradictory. But, the consistency of these 
is a consequence of the fact that the mutual hindrance leads to 
the smaller average speed $v_c = q$ in spite of the higher hopping 
probability $u(x) = Q$.

It should be noted \cite{nds} that (\ref{cri}) is also obtained
by the intersection point of the line $F=v_c \rho$ and the curve 
\cite{css}
\begin{equation}
 F=\frac12(1- \sqrt{1-4Q\rho(1-\rho)})
\label{asep}
\end{equation}
that describes the fundamental diagram of the ASEP with parallel 
updating. Also note that the limits $L \to \infty$ and $f \to 0$ do 
not commute\cite{LEC}. If we take $f \to 0$ before $L \to \infty$, 
then we apparently have (\ref{asep}), which corresponds to the 
situations in our numerical simulations. On the other hand, if
we take $f \to 0$ after $L \to \infty$, then we have the thick 
curve in Fig.3. Thus, in the latter case, we have proved 
that the anomalous variation of the average velocity with the
density disappears.

\begin{figure}[ht]
\begin{center}
\includegraphics[width=0.8\columnwidth]{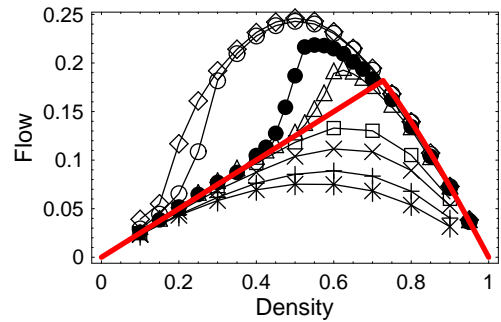}
\caption{The thick curve (with no decorating symbols) represents the 
fundamental diagram of the ATM corresponding to $f\to 0$ {\em after}
taking the thermodynamic limit $L\to\infty$. 
Note that these two limits do not commute.
The simulation data corresponding 
to $f=0.0005 ({\Diamond})$, $0.001 (\circ$), $0.005 (\bullet)$, 
$0.01 ({\bigtriangleup})$, $0.05 ({\Box})$, $0.10 (\times)$, 
$0.25 (+)$, $0.50 (\ast)$.  
}
\end{center}
\label{thermo}
\end{figure}
\section{Phases of ATM with open boundary conditions}\label{sec4}

So far we have considered the ATM with only periodic boundary 
conditions. However, for ant-trails the open boundary conditions 
are more realistic. Therefore, in this section we study the  
phases and phase transitions in the ATM with open boundary conditions. 

Suppose $\alpha$ and $\beta$ denote the probabilities of incoming 
and outgoing particles at the open boundaries per unit time.
The phase diagram of the ASEP in the $\alpha-\beta$-plane has been 
investigated exhaustively \cite{schutz}. In this section we report 
the effects of varying the pheromone evaporation probability $f$ 
on the phase diagram of the ASEP with parallel updating and open 
boundary conditions.

\begin{figure}[ht]
\begin{center}
\includegraphics[width=0.75\columnwidth]{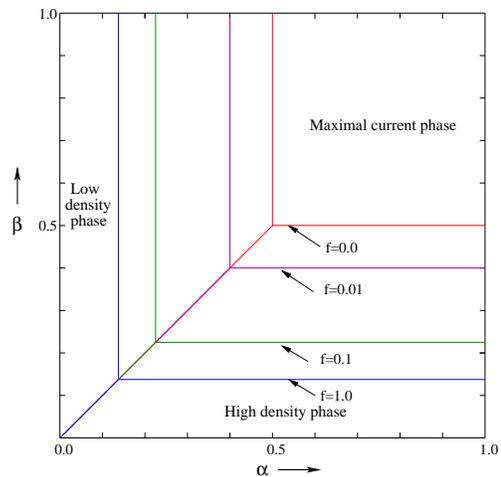}
\caption{The phase diagram of the ATM with open boundary conditions 
in the $\alpha-\beta$-plane for several different values of the 
pheromone evaporation probability $f$ ($0 \leq f \leq 1$). The values 
of the hopping parameters are $Q = 0.75, q = 0.25$.
}
\end{center}
\label{obcphase}
\end{figure}

Just as in the case of the ASEP, we found three different phases, 
namely, the high-density phase, the low-density phase and the 
maximal current phase of the ATM for all $f$ (see Fig.4). 
The line separating the low-density phase and the maximal current 
phase is given by $\alpha_c(f)$ whereas that corresponding to the 
separation between the high-density phase and the maximal current 
phase is given by $\beta_c(f)$.

\begin{figure}[ht]
\begin{center}
\includegraphics[width=0.65\columnwidth,angle=-90]{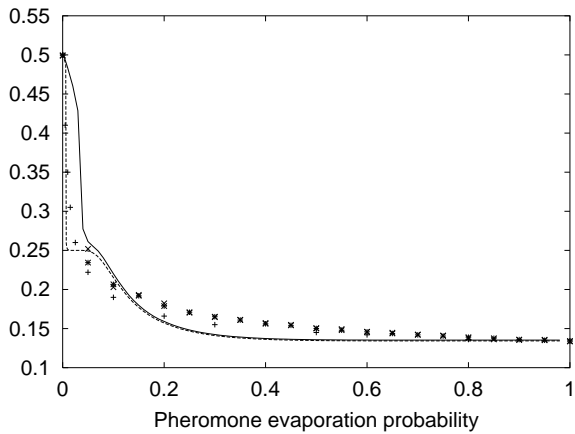}
\caption{The critical rates, $\alpha_c$, (denoted by $+$),  
which have been obtained from computer simulations of the ATM 
with open boundary conditions, are plotted against the pheromone 
evaporation probability $f$. The discrete data points denoted 
by the symbols $\times$ and $\ast$ denote the average speeds 
$v_{1/2}(f)$ obtained from computer simulations of the same 
model but with periodic boundary conditions for system sizes 
$L = 100$ and $L = 500$, respectively. The corresponding 
theoretical predictions for $v_{1/2}(f)$ of our ZRP-based 
theory of ATM are plotted for the system sizes $L = 100$
(the continuous curve) and $L = 500$ (the dashed curve), 
respectively. The values of the hopping parameters are 
$Q = 0.75, q = 0.25$.
}
\end{center}
\label{critical}
\end{figure}

The variation of the critical point $\alpha_c(f) = \beta_c(f)$ 
with $f$ is shown in Fig.4. In order to understand 
this observation let us first examine the limiting balues at 
$f = 0$ and $f = 1$. It is well known \cite{rajewsky} that for 
ASEP with hopping probability $q_{\text{eff}}$, 
\begin{equation} 
\alpha_c = \beta_c = 1 - \sqrt{1-q_{\text{eff}}}. 
\label{alphacasep}
\end{equation}
Therefore, we must have 
\begin{eqnarray} 
\alpha_c(f=0) = \beta_c(f=0) = 1 - \sqrt{1-Q} \nonumber \\
\alpha_c(f=1) = \beta_c(f=1) = 1 - \sqrt{1-q}~~; 
\end{eqnarray} 
these are consistent with the numerical data $\alpha_c(f=0) = 0.5$ 
and $\alpha_c(f=1) = 0.133$ for $Q = 0.75, q = 0.25$ shown 
in Fig.4. 

Next we try to understand the detailed variation of 
$\alpha_c = \beta_c$ with $f$ by utilizing the relation between 
the ATM model with periodic and open boundary conditions.
Note that, in the maximal current phase in the ASEP with open 
boundary conditions and parallel updating \cite{rajewsky} 
the current is given by 
\begin{equation}
J = \frac{1-\sqrt{1-q_{\text{eff}}}}{2} 
\label{mcflux}
\end{equation}
(in ASEP with parallel updating) while the corresponding bulk density 
is given by \cite{rajewsky} 
\begin{equation}
\rho(x=L/2) = 1/2 
\label{mcrho}
\end{equation}
and, consequently, 
the corresponding average speed should be 
\begin{equation}
v = 1-\sqrt{1-q_{\text{eff}}} = \alpha_c = \beta_c.
\label{mcvel}
\end{equation}
In order to check the validity of this argument, we have computed 
the average speed $v_{1/2}$ corresponding to $\rho = 1/2$ by (a) 
computer simulation of the ATM with periodic boundary 
conditions and (b) using the ZRP-based theory, mentioned 
in the preceeding section. All these data are plotted in  
Fig.5. The simulation data for $v_{1/2}(f)$ are 
in good agreement with the simulation data for $\alpha_c$. But, 
there are significant differences between these data and 
$v_{1/2}(f)$ obtained from our ZRP-based theory in the small $f$ 
regime. We believe that this discrepancy arises from the boundary 
effect. If periodic boundary conditions are imposed, at low 
densities, the leading ant in the loose cluster can smell the 
pheromone which is dropped by the last ant in the same cluster. 
However, this effect would disappear when the periodic boundary 
conditions are replaced by open boundary conditions.

Since we have numerically estimated $\alpha_c$ in the ATM as a 
function of $f$ by carrying out computer simulations, we utilized 
the relation (\ref{alphacasep}) to get the $f$-dependence of the 
effective hopping probability $q_{\rm eff}(f)$ in the ATM (see 
Fig.6). The two limiting values 
$\lim_{f \to 0} q_{\rm eff}(f) = Q$ and
$\lim_{f \to 1} q_{\rm eff}(f) = q$ as well as the nature of the 
variation of $q_{\rm eff}$ with $f$ are fully consistent with one's 
expectation based of physical arguments mentioned earlier in this 
paper.

\begin{figure}[ht]
\begin{center}
\includegraphics[width=0.65\columnwidth,angle=-90]{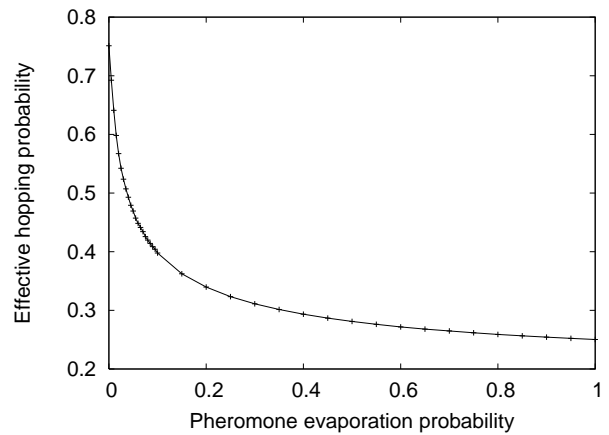}
\caption{$f$-dependence of the effective hopping probability, 
extracted by using the equation (\ref{alphacasep}) from the numerical 
data obtained by computer simulation of the ATM with open boundary 
conditions. 
}
\end{center}
\label{qeff}
\end{figure}

\section{Concluding discussions}\label{sec5}

The close similarities between phase transitions in non-living 
systems and that of foraging behaviour on the ant-trail was pointed 
out by Beckman et al.\cite{beckman}. However, their study was 
concerned with the nature of foraging, namely, the possibility 
of a transition from a disordered foraging behaviour (i.e., 
foraging without a pheromone trail) to ordered foraging (i.e., 
trail-based foraging). In contrast to this problem of the formation 
of the trail pattern \cite{phasetr}, we have studied the phases 
and phase transitions exhibited by the system when the collective 
traffic-like movement of ants take place on an existing trail.

The dynamical phases and non-equilibrium phase transitions in 
systems of interacting self-propelled particles have been among 
the most challenging problems of investigation in non-equilibrium 
statistical mechanics. In contrast to equilibrium systems, these 
intrinsically non-equilibrium systems exhibit phase transitions 
even in one-dimensions with only short range interactions. 
In this paper we have studied the phases of a one-dimensional 
model motivated by the collective traffic-like movements of ants 
on an existing trail \cite{cgns,nds}. In this model the ants are 
represented by self-propelled particles which, in addition to the 
hard-core repulsion, interact indirectly via pheromone. The model is, 
thus, characterized by two coupled dynamical variables, representing 
the ants and the pheromone. 

In our earlier works \cite{cgns,nds} we had shown that the 
homogeneous mean-field approximations cannot capture the non-monotonic 
variation of the average speed with the density of the ants. Even 
the loose-cluster approximation that we developed\cite{nds} 
could account for the simulation data with limited accuracy. 
In this paper we have reported our new quantitative results on this 
model with periodic boundary conditions; these results have been 
derived by utilizing the analogy with ZRP. Moreover we have shown 
that there is a phase transition in the thermodynamic limit in this 
model, albeit in a special limit $f \to 0$. 

In our earlier published works \cite{cgns,nds} we imposed periodic 
boundary conditions. However, in order to capture the real ant-trails 
in nature imposition of open boundary conditions seems more appropriate.
In equilibrium statistical mechanics, the boundary conditions do not 
play any role in the phase transitions which, strictly speaking, 
take place only in the thermodynamic limit. In contrast, 
boundary-induced phase transition \cite{krug} is a well-known phenomenon 
in non-equilibrium statistical mechanics. The phase diagrams of the 
ASEP with open boundary conditions is now well established. Since 
the ATM can be regarded as an extension of the ASEP, our investigation 
of the phase diagram of the ATM reported here illuminates the effects 
of varying the pheromone evaporation rate $f$ on the phase diagram of 
the ASEP. In this paper we have focussed attention on phase transitions 
in an idealized single-lane ATM. The phase transitions in a more 
elaborate two-lane ATM will be reported later \cite{johnetal}.
It would be interesting to test the predictions of the model, 
particularly the non-monotonic variation of average speed with 
density, by repeating experiments of the type reported in ref.10 
to get sufficiently accurate data for real ants. 

\noindent{\bf Acknowledgements} We thank Martin Evans for drawing 
our attention to the analogy between ATM and ZRP. We also thank 
Joachim Krug, Vladislav Popkov and Martin Evans for enlightening 
discussions.


\end{document}